\documentclass[preprint2]{aastex}

\def\msol{M_\odot}
\def\mjup{M_{\rm J}}
\def\msolyr{M_\odot \rm {yr}^{-1}}
\def\msun{M_\odot}
\def\rsun{R_\odot}

\def\te{T_{\mathrm{eff}}}
\def\mdot{\dot{M}}
\def\dt{\Delta t}
\def\minit{m_{\rm i}}
\def\be{\begin{equation}}
\def\ee{\end{equation}}
\def\mjup{M_\mathrm{{Jup}}}

\def\lsun{L_\odot}

\def\simgr{\,\hbox{\hbox{$ > $}\kern -0.8em \lower 1.0ex\hbox{$\sim$}}\,}
\def\simle{\,\hbox{\hbox{$ < $}\kern -0.8em \lower 1.0ex\hbox{$\sim$}}\,}

\shorttitle{Episodic accretion}
\shortauthors{Baraffe et al.}

\begin{document}

\title{Episodic accretion at early stages of evolution of low mass stars and brown dwarfs: a solution for the observed luminosity spread in HR diagrams?}


\author{I. Baraffe\altaffilmark{1} and G. Chabrier\altaffilmark{1}}
\affil{\'Ecole Normale Sup\'erieure, Lyon, CRAL (UMR CNRS 5574), Universit\'e de Lyon, France}
\email{ibaraffe,chabrier@ens-lyon.fr}
\and
\author{J. Gallardo\altaffilmark{2}}
\affil{Observatorio Astron\'omico Cerro Cal\'an, Departamento de Astronom\'{\i}a, Universidad de Chile, Casilla 36-D
              Santiago, Chile}
               \email{gallardo@das.uchile.cl}

\begin{abstract}
We present  evolutionary models for young low mass stars
and brown dwarfs taking into account episodic phases of accretion at 
early stages of the evolution, a scenario supported by recent large surveys of embedded protostars. An evolution including short episodes of vigorous
accretion ($\mdot \ge 10^{-4} \msolyr$) followed by longer quiescent phases
($\mdot < 10^{-6} \msolyr$)
can explain the observed luminosity spread in HR diagrams 
of star forming regions at ages of a few Myr, for objects ranging from a few Jupiter masses to a few tenths of a solar mass. The gravitational contraction of
these accreting objects strongly departs from the standard Hayashi track at constant $\te$.
%A significant luminosity spread is obtained  if the non steady accretion phase starts on central proto-objects of different masses,from a few $\mjup$ to a few tenth of $\msun$. 
The best agreement with the observed luminosity scatter is obtained if most of the accretion shock energy is radiated away.
%, consistent with prior cooling through a disk. 
The obtained luminosity spread at 1 Myr in the HR diagram is equivalent to what can be misinterpreted as a $\sim$ 10 Myr age spread for non-accreting objects.
We also predict a significant spread in radius at a given $\te$,
as suggested by recent observations. 
These calculations bear important consequences on our understanding of star formation and early stages of evolution and on the determination of the IMF for young ($\le$ a few   Myr) clusters. 
Our results also show that the concept of a stellar birthline for low-mass objects has no valid support.
\end{abstract}

\keywords{stars: formation --- stars: low-mass, brown dwarfs --- accretion, accretion disks}

\section{Introduction} \label{intro}

The significant luminosity spread observed in Hertzsprung-Russell (HR) diagrams of star forming regions (SFRs) and young clusters  is a well known feature, which has been confirmed with the improvement of observational techniques
(see Hillenbrand 2008 and references therein). 
%Despite many efforts devoted to better observational data, this spread persists, raising the question 
Whether this luminosity spread arises from a physical process, observational uncertainties  or reveals a significant age spread is a crucial question, with important consequences on our understanding of star formation
(Hartmann 2001). This hypothetical age spread is used as an argument in favour
 of a slow star formation process, in conflict with other observational constraints and our current
understanding of star formation
 (Hartman 2001; Ballesteros-Paredes \& Hartmann 2007).
Motivated by this 
%puzzling feature and this 
controversy,
we have conducted a systematic analysis of how accretion affects the evolution of young low mass stars (LMS) and brown dwarfs (BDs) in order to explore the
sensitivity of evolutionary tracks to the early accretion history. 
In this Letter, we present the first consistent evolutionary models for young LMS
and BDs taking into account non-steady accretion phases at very early stages of the evolution. 
We show that this scenario can explain at least partly the observed luminosity spread in HR diagrams, without invoking an age spread. 
%The details of the calculations will be presented in a forthcoming paper (Gallardo et al. 2009, in prep.). 

Current observational analysis of embedded protostars strongly suggest that accretion onto forming stars must be transient, with very large fluctuations (Dunham et al. 2008; Enoch et al. 2009; Evans et al. 2009). Enoch et al. (2009) find in three clouds a large population of low luminosity class I sources 
that aggravate the well known "luminosity problem" (Kenyon et al. 1990). 
Long quiescent phases of accretion ($\mdot \simle 10^{-6} \msolyr$) interrupted by short
episodes of high accretion ($\mdot \simgr 10^{-5} \msolyr$) provide a consistent picture explaining both the large population of
low luminosity class I sources and the small fraction of very luminous sources.
Enoch et al (2009) also rule out drastic changes in the accretion rates from class 0 to class I,
and in particular the  standard picture of short ($\sim 10^4$ yr) class 0 duration\footnote{
Standard estimates 
for ages and rates are $\sim 10^4$ yr and $\mdot\sim 10^{-5} \msolyr$ 
for class 0 and $\sim 10^5$ yr and $\mdot\sim 10^{-6} \msolyr$ for 
class I sources.}.
Evans et al. (2009)
suggest that a star could assemble half its mass during a few episodes of high accretion,
occurring throughout about 7\% of the class I lifetime. The idea of
non steady, time varying accretion rates is
not new, since for decades FU Ori objects
have been providing evidence for short episodes of rapid accretion at early stages of
evolution, with rates much larger than the aforementioned typical class-0 infall rates for low mass objects (Kenyon \& Hartmann 1995).

 The calculations presented in this Letter are a theoretical formulation of such an evolution including phases of episodic accretion for proto- or young LMS and BDs.
%with these recent observations and interpretations. 
In \S 2, we briefly summarize the evolutionary models and the treatment of
accretion; details will be presented in a forthcoming paper (Gallardo et al. 2009). Results and comparison with observations are presented in \S 3, followed by discussion and
conclusion in \S 4. 

\begin{table*}
\begin{center}
\caption{Parameters of the evolutionary sequences 
shown in Fig. \ref{hrd1}. $\minit$ and $m_{\rm f}$ are
respectively the initial (in $\mjup$)
and final  (in $\msol$) masses, $T_{\rm eff1}$ and $L_1$
the effective temperature and log of the luminosity (in units of $\lsun$) respectively 
at 1 Myr, $\te^{\rm BCAH98}$ and $L^{\rm BCAH98}$ the values 
predicted by the BCAH98 (non accreting) models for the mass  $m_{\rm f}$ at 1 Myr,
$\mdot$ the accretion rate  (in $\msolyr$) applied during a time  $\dt$ (in yr) and
$\alpha$ the fraction of accretion energy absorbed by the
object (assuming $\epsilon$=1/2, see text).
\label{table1}}
\begin{tabular}{cccccccccc}
\tableline\tableline
case & $\minit$ & $m_{\rm f}$ & $T_{\rm eff1}$ &$L_1$ & $\te^{\rm BCAH98}$ & $L^{\rm BCAH98}$ & $\mdot$ & $\dt$  & $\alpha$  \\
 \tableline
A &1 & 0.05 & 2860  &-2.15 & 2844 & -1.73 & 10$^{-5}$ &  5 10$^3$ & 0 \\
B &1 & 0.1 & 3157 &-1.63 & 3001 & -1.16 & 10$^{-5}$ &  10$^4$ & 0 \\
C &1 & 0.2 & 3320 &-1.27 & 3193 & -0.69 & 10$^{-5}$ & 2 10$^4$ & 0 \\
D &1 & 0.5 & 3756 & -0.75& 3426 & -0.28 & 5 10$^{-5}$ &  10$^4$ & 0 \\
E &1 & 0.1 & 3000 & -1.16 &  3001 & -1.16 & 10$^{-5}$ &  10$^4$ & 0.2 \\
F &1 & 0.1 & 2998 & -1.16 & 3001 & -1.16 & 10$^{-5}$ &  10$^4$ & 1 \\
G & 5 & 0.1 & 3051 & -1.53 & 3001 & -1.16 & 10$^{-5}$ &  10$^4$ & 0 \\
H &10 & 0.21 & 3221 & -1.20 & 3201 & -0.67 & 10$^{-5}$ & 2 10$^4$ & 0 \\
I & 50 & 0.55 & 3571 & -0.59 & 3468 & -0.22 & 10$^{-5}$ &  5 10$^4$ & 0 \\
J & 50 & 1.05 & 4102 & -0.32 & 3840 & 0.17 & 5 10$^{-5}$ & 2 10$^4$ & 0 \\
K &100 & 1.1 & 3961 & -0.03 & 3870 & 0.20 & 5 10$^{-5}$ & 2 10$^4$ & 0 \\
L &100 & 1.85 & 4677 & 0.31 & 4377 & 0.60 & 5 10$^{-5}$ &  3.5 10$^4$ & 0 \\
\tableline
\end{tabular}
\end{center}
\end{table*}

\section{Evolutionary models and treatment of accretion}\label {section_models}

The evolutionary calculations for LMS and BDs
are based on the Lyon stellar evolution code with input physics
described in Chabrier \& Baraffe (1997) and Baraffe et al. (1998). The treatment of accretion
 is based on a simplified 1D approach, adopting similar assumptions and simplifications
as Hartmann et al. (1997) and Siess et al. (1997). 
We assume that accretion onto the central object rapidly proceeds non-spherically,
affecting only a small fraction $\delta$ of the contracting object's surface. 
%$S_\star=4\pi R^2$. 
The object can thus
freely radiate its energy over most of its photosphere (see Hartmann et al. 1997 and references therein). 
%Such an assumption of accretion covering a small fraction of the stellar surface is supported by various observations of the optical continuum emission of accreting T-Tauri stars (see Hartmann et al. 1997 and references therein). 
The accreting material brings, per unit mass, a gravitational energy $-GM/R$
and an internal energy $+\epsilon GM/R$, i.e. an energy rate:
%assuming it gains energy at a rate $\alpha L_{\rm acc}$, with: 
\be
%L_{\rm acc} = a\pi R^2 {\cal F}_{\rm acc}= \epsilon { G M \mdot \over R} 
\frac{dE_{\rm acc}}{dt}=( \epsilon -1) { G M \mdot \over R}
\ee
The value of $\epsilon$ depends on the details of the accretion process, with $\epsilon \le{1}$ for gravitationally bound material and
 $\epsilon \le{1 \over 2}$ if gas accretes from a thin disk at the object's equator (Hartmann et al. 1997). We denote $\alpha$ ($\alpha \le 1$)
the fraction of
accreting internal energy absorbed by the proto-star/brown dwarf, which thus contributes to its heat content. The total additional energy rate gained by the accreting object and the accreting luminosity radiated away thus read, respectively, for $\delta \ll 1$ (see e.g. Hartmann et al. 1997):
 \be
L_{\rm add} =  \alpha \epsilon { G M \mdot \over R};\,\, L_{\rm acc} 
%= 4\pi R^2\delta {\cal F}_{acc}
=\epsilon(1-\alpha) { G M \mdot \over R}
\ee
%whereas the fraction $\epsilon(1-\alpha) L_{\rm acc}$ is radiated away.
The case $\alpha \ll 1$ corresponds to accreted matter
 arriving on the object's surface with a lower specific entropy than the object's one. 

%\begin{figure}[h!]
\begin{figure}
\begin{center}
%\epsscale{1.1}
%\plotone{hrd1.ps}
%\includegraphics[height=8cm]{hrd1.ps}
\includegraphics[height=11cm]{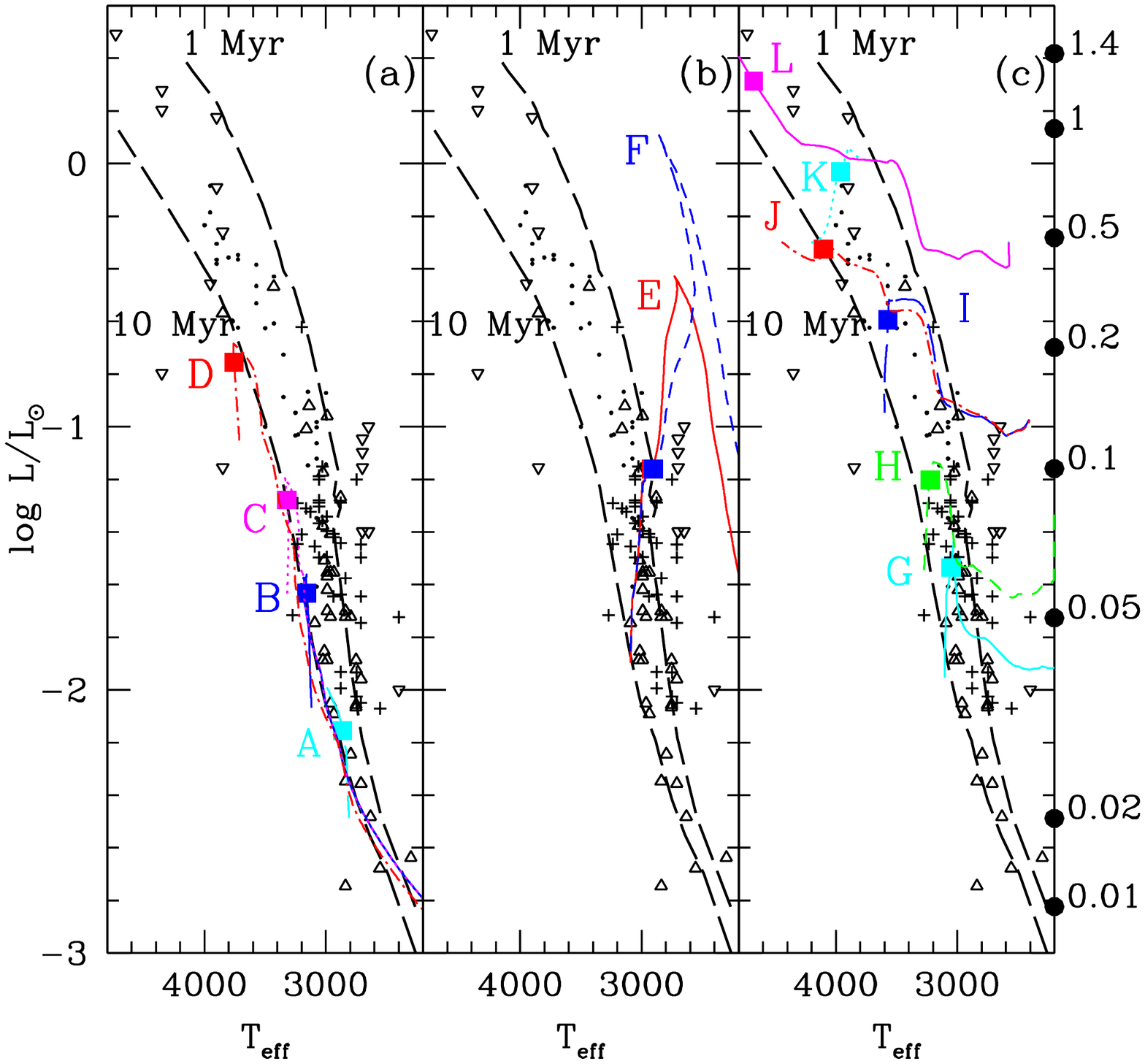}
\caption{Effect of accretion on the evolution in a HRD for objects with different initial
 masses $m_{\rm i}$. The two long dashed (black) curves are the 1 Myr and 10 Myr isochrones of Baraffe et al. (1998)
for non accreting models.
The solid (black) circles on the rightmost Y-axis give the luminosity of non-accreting models at 1 Myr for the indicated masses (in $\msol$).
%The full circles on these curves indicate the position of 0.01, 0.02, 0.05, 0.1, 0.2, 0.5 and 1 $\msol$.
The coloured curves show calculations for various accretion
rates described in Table \ref{table1}. The solid squares on each curve gives the position at 1 Myr.
{\bf (a)}: Initial mass $m_{\rm i} = 1 \mjup$ and $\alpha$=0.
%Case A (long-dash cyan) - case B (solid blue) -  case C (dot magenta)  - case D (dash-dot red).
{\bf (b)}:  initial mass $m_{\rm i} = 1 \mjup$ and $\alpha \ne 0$.
%Case E (solid red) and case F (dash blue).
Note that both sequences reach the same location at 1 Myr.
{\bf (c)}: Initial masses $\minit > 1 \mjup$.
%Case G (solid cyan) -  case H (dash green) -  case I (long dash blue)
%- case J (dash-dot red) -
%case K (dot cyan) - case L (solid magenta).
Other symbols are observations
in star forming regions ({\small $\nabla$} Gatti et al. 2006;  {\large \bf .} Gatti et al. 2008; $+$ Peterson et al. 2008; {\small $\Delta$} Muzerolle et al. 2005).}
\label{hrd1}
\end{center}
\end{figure}

\section{Results} 

\subsection{Exploring accretion rates} \label{subsection_test}

We have conducted evolutionary calculations taking into account the effect of accretion on proto low mass objects for a wide range of initial proto-star/BD masses from $10^{-3}\msol$ ($\sim$ 1 $\mjup$) to 0.1 $\msun$, with arbitrary large initial radii between $\sim 1 \rsun$ and $\sim 4 \rsun$.
The results are compared to recent
surveys of LMS and BDs between $\sim$ 0.01 $\msol$ and 1 $\msol$ in various SFRs, with characteristic ages
of a few Myr, namely
Taurus and Chamaeleon I (Muzerolle et al. 2005), $\rho$ Ophiucus (Gatti et al. 2006),
Orion molecular cloud (Peterson et al. 2008), $\sigma$-Orionis (Gatti et al. 2008).
Given the loose constraints on accretion rates at early times, during the class 0 and class I
embedded phases, we have explored a wide range
of mass accretion rates and time dependences, and considered different values of $\alpha$. We have considered
(1) constant rates, (2) exponentially time-decreasing rates
and (3) rates obeying the empirical mass dependence $\mdot \propto M^2$  observed in young clusters (Mohanty et al. 2005, Herczeg \& Hillenbrand 2008). 
The strong constraint of our calculations is to be consistent with the observed rate determinations at an age of a few Myr.
 
We find that assuming initial accretion rates $\sim 10^{-6} \msolyr$, as traditionally used in
protostellar evolutionary models
(Myers et al. 1998; Young \& Evans 2005), 
%characteristic of  the standard collapse mass infall rate ${\dot M}\simeq c_{\rm s}^3/G$ (Shu 1977),
produces too small a luminosity scatter in HR diagrams after a few Myr
to explain the observed spread. More severe effects are obtained when adopting higher accretion rates, 
$\mdot \sim (1-5)\times 10^{-5} \, \msolyr$, during the first
few $10^3$ to $10^4$ years of evolution, depending on the mass.
Fig. \ref{hrd1} shows the effect of such high early accretion rates on the evolution
of low-mass objects for different initial masses $m_i$. Evolution proceeds as time increases 
from the right to the left part of the HR diagram for a given track. The tracks of the accreting 
objects are displayed
up to an age of 10 Myr with the locations at 1 Myr indicated by the solid squares.
% on objects of initial masses ranging between 1 $\mjup$ and 0.01 $\msol$, 
%for various accretion rates $10^{-5} \msolyr \le \mdot \le 5 \,10^{-5} \msolyr$ applied during a time
% 10$^4 {\rm yr} \le \dt \le 5 \,10^4$ yr.
The evolutionary sequences start from
large initial radii and thus with short
thermal timescales, $\tau_{\rm KH} \approx {G M^2 \over RL} \sim 10^3$ yr, of the order of or less than the accretion timescale $\tau_{acc}\sim M/{\dot M}$.
Consequently, 
variations of the initial radius  by a factor 2-3
have no significant effect on the overall evolution, 
%between $\sim 0.5\, \rsun$ and a few $\rsun$  
barely changing  the final location at 1 Myr.
%since in all cases the thermal timescale remains very short {\bf estimate ?, you mean $<<10^6$ ?}, 
%and thus our conclusions. 
The parameters of the various sequences are given in Table \ref{table1}.

\begin{figure}[h!]
\begin{center}
\includegraphics[height=10cm]{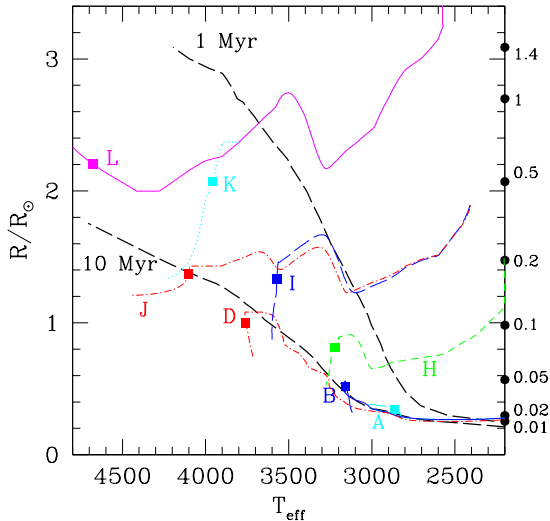}
%\epsscale{0.9}
%\plotone{radius.ps}
\caption{Evolution of the radius versus $\te$ for accreting objects of different
initial masses. The colored curves correspond to different
accreting sequences described in Table \ref{table1}.
The other curves and symbols are the same as in Fig. \ref{hrd1}. The solid (black) c
ircles on the rightmost Y-axis give the radius
of non-accreting models at 1 Myr for the indicated masses (in $\msol$).}
\label{radius}
\end{center}
\end{figure}

For $\alpha$=0, i.e. if all the accreting energy is radiated away, the aforementioned high accretion rates yield
significantly smaller radii than the ones of non accreting objects of same mass
and age.
%, as mentioned in \S \ref{section_models}. 
This stems from the fact that, as mass builds up, the object's  thermal timescale $\tau_{\rm KH}$
%, $\tau_{\rm KH} \approx {G M^2 \over RL}$, 
rapidly increases,
%with $\tau_{\rm KH}\sim 10^7$ yr after a few 10$^4$ yr,
 and becomes much longer than the accreting timescale
 $\tau_{acc}$,
%\sim M/{\dot M}$, 
so that rapid increase of gravitational energy is the only possibility to adjust to the accreting energy flow.
For example, the case B sequence (solid blue line in Figs. \ref{hrd1}a and \ref{radius}) starts its evolution
with $m_i=1\,\mjup$ and $\tau_{\rm KH} \sim 10^3$ yr. At 10$^4$ yr, when accretion is arbitrarily
stopped, the object has a mass 0.1 $\msol$, a radius $\sim 0.5 \, \rsun$, a luminosity
$L \sim 2.5 \times 10^{-2} \,L_\odot$ and a thermal timescale  $\tau_{\rm KH} \sim 2 \times10^7$ yr. Similar results are obtained for all the other sequences starting from larger initial masses. 
%the sequence with $m_{\rm i} = 0.1 \msol$, initial radius 
%3.6 $\rsun$ and $\mdot = 5 \,10^{-5} \msolyr$ during $\dt = 2 \,10^4$ yr (dot - long dash cyan line in Figs. \ref{hrd1}-\ref{radius}). At an age of 
%$2 \,10^4$ yr, the mass is 1.1 $\msol$, radius $\sim 2.2 \rsun$, $L \sim \lsun$
%and thus $\tau_{\rm KH} \sim 10^7$ yr.  
The radius of these accreting objects
is thus already smaller after a few $10^4$ yr than the one of the non-accreting counterparts of
same mass at an age of 1 Myr (see Fig. \ref{radius}). The objects will eventually slowly
contract toward their location on the HR diagram at 1 Myr after the high accretion phase has been stopped, looking much
fainter than the non-accreting 1 Myr old objects (see Fig. \ref{hrd1}).
The spread in luminosity obtained in Fig. \ref{hrd1} thus essentially reflects
a spread in radius, as illustrated in Fig. \ref{radius}.  Note that we  
find similar results if we apply smaller, non-zero accretion rates, $\mdot \simle 10^{-6} \msolyr$, after the strong accretion phase,  while fulfilling the condition to recover typical  observed rates at
$\sim$ 1 Myr. 

For the case $\alpha \ne 0$, i.e if some fraction of the accretion energy contributes to the proto-star/BD thermal content,
the spread at 1 Myr is smaller than for the $\alpha$=0 case. Assuming  $\epsilon$=1/2 in Eq. (2),
characteristic of accretion from a thin disk (see \S 2),
we find that for $\alpha \simle 0.2$, the contraction of the
structure due to mass accretion is partly compensated by this energy input, yielding a less drastic
contraction. The objects are still slightly fainter 
at $\sim$ 1 Myr  than their non-accreting counterparts, 
but the effect is not as significant as in the $\alpha$=0 case.
For $\alpha \simgr 0.2$,
the effect of the extra accreting energy contribution to the protostar heat content becomes dominant and the evolution now mostly proceeds
at higher luminosity (see Fig. \ref{hrd1}b) and larger radius than for non-accreting objects. This in turn implies
{\it shorter} thermal timescales, so that once high accretion rates are 
stopped or significantly decreased to recover the observed values at $\sim$  1 Myr,
the object quickly contracts and reaches a position at 1 Myr in the HR diagram very close to
the non accreting location. In order to maintain a high luminosity
and radius after $\sim$ 1 Myr, {\it i.e} a location in the HR diagram {\it above} the 1 Myr non accreting isochrone, too large accretion rates  (and large values of $\alpha$) have to be maintained, inconsistent with observations of class II objects at this age. 

\subsection{Episodic accretion}

\begin{figure}[h!]
\begin{center}
\includegraphics[height=11cm]{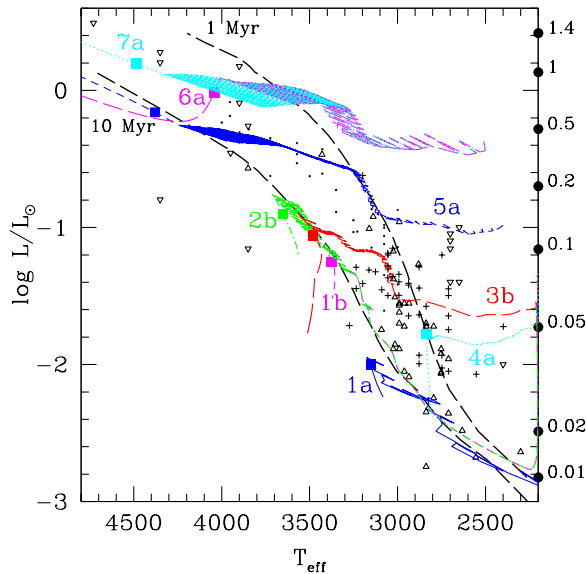}
%\epsscale{1.1}
%\plotone{burst.ps}
\caption{Evolutionary sequences in a HRD with episodic accretion.
The labels close to each colored curve correspond to the cases described in Table \ref{table2}.
%The label $a$ or $b$ indicates
%if the bursts start already at the beginning of evolution (case $a$) or after a first phase of constant accretion ($\mdot = 10^{-5} \msolyr$ during $\dt = 10^{4}$ yr) (case $b$).
%Case 1$a$ (solid blue) -  case 1$b$ (dash magenta) -
%case 2$b$ (dash-dot green) -  case 3$b$ (long dash red) -
%case 4$a$ (dot cyan) -
%dash (blue): case 5$a$; dot- long dash (cyan): case 6$a$.
Symbols and other curves are the same as in Fig. \ref{hrd1}. }
\label{burst}
\end{center}
\end{figure}

The exploratory analysis presented in the previous section demonstrates that the
early accretion history still significantly affects the evolution of contracting objects after a few Myr.
 In this section, we show that
the effects of accretion are similar, whether
the object accretes {\it continuously} an amount of mass $\Delta M$ at a constant rate 
$\mdot \sim 10^{-5} \msolyr$ during $\dt \sim 10^4$ yr, as examined in \S3.1, or whether it accretes the same amount of mass
during a {\it succession of short episodes} of high accretion rates, 
$\mdot \ge 10^{-4} \msolyr$, interrupted by longer quiescent phases.  
As mentioned in \S \ref{intro}, episodic accretion seems to provide the most consistent
explanation for current observations of protostars, while a short ($\sim 10^4$ yr) accretion phase
based on previous estimates of class 0 lifetimes, with accretion rates significantly 
decreasing during  the subsequent class I phase, seems to 
be ruled out (Enoch et al. 2009; Evans et al. 2009). Several theoretical scenarios
have been suggested to explain phases of non-steady accretion (Kenyon \& Hartmann 1995; Vorobyov \& Basu 2005; Zhu et al. 2008; Tassis \& Muschovias 2005). 
%Some models
%suggest constant infall from the envelope onto the disk ({\it e.g} Kenyon \& Hartmann 1995) and episodic accretion from the disk onto the protostar, due to gravitational
%instabilities (Vorobyov \& Basu 2005) or to a combination of gravitational
%and magnetorotational instabilities (Zhu et al. 2008). On the other hand, Tassis \& Muschovias (2005)
%suggest  magnetically controlled spasmodic infall onto the  disk, yielding
%strong variations of accretion onto the central object. 
We adopt here a burst mode as suggested by Vorobyov \& Basu (2005)
to explore the effect of episodic accretion on the structure of
proto-stars/BDs. 
We assume an arbitrary number of bursts, $N_{\rm burst}$, with typical accretion rate $\mdot_{\rm burst} \ge 10^{-4} \msolyr$ and duration $\dt_{\rm burst} \sim 100$ yr, interrupted by quiescent phases
of duration $\dt_{\rm quiet}$= 1000-5000 yr (see Vorobyov \& Basu 2005). During the quiescent phases,
we adopt $\mdot$=0 as a simplification\footnote{Adopting rates  $\mdot < 10^{-6}$ provides the same qualitative results and
does not affect our conclusions.}. We have explored two possibilities for the beginning
of the burst phase (cases {\it a} and  {\it b} described in Table \ref{table2}).
%either it starts from the very beginning of the evolution (case $a$) or it starts
%after an early phase of constant accretion ($\mdot = 10^{-5} \msolyr$ during
%10$^4$ yr), as suggested
% by Vorobyov \& Basu (2005) (case $b$).
The results are displayed in Fig. \ref{burst}.
The burst phases are characterised by evolutionary tracks which follow an erratic behavior with abrupt variations of $L$ and $\te$.
Our calculations show that, depending
on $\minit$, $N_{\rm burst}$ and $\mdot_{\rm burst}$,
it is possible to populate the region in the HR diagram after $\sim$ 1 Myr or less between the (non-accreting) 1 Myr and 10 Myr isochrones, 
producing a natural spread in luminosity. For values of $\mdot_{\rm burst} \ge 10^{-4} \msolyr$, the evolution is severely affected while for smaller rates,
 the structure is only moderately affected and the object has time to relax once episodic accretion stops,
 having properties
at $\sim$ 1 Myr similar to the non accreting counterpart
of same mass and age. This is illustrated by 
case 4a (see Table \ref{table2}) in Fig. \ref{burst} (cyan dotted curve), with $\mdot_{\rm burst} = 10^{-5} \msolyr$. Similar evolutionary properties
are obtained adopting case $a$ or $b$ for the beginning of the burst phase.
The duration of the quiescent phase $\dt_{\rm quiet}$ is found to be inconsequential and can be increased from 10$^3$ yr to $\sim$ 10$^4$ yr
without significant effects. This reflects the fact that the thermal timescale of
the accreting object rapidly exceeds values $\gg 10^4$ yr. The contracting object has thus no time to relax
to a larger radius state for its new mass if  quiescent phases last less than 10$^4$ yr.
For typically 10 to 100 burst episodes, this means that episodic accretion can last a few 10$^5$
yr, in agreement with recent revised estimates of class 0 and class I lifetimes
(Enoch et al. 2009;  Evans et al. 2009). 

\begin{table*}
\begin{center}
\caption{Parameters of the evolutionary sequences assuming episodic accretion and
shown in Fig. \ref{burst}.
$\minit$, $m_{\rm f}$, $T_{\rm eff1}$, $L_1$, $\te^{\rm BCAH98}$ and $L^{\rm BCAH98}$ have the same meaning as in Table 1.
%are respectively the initial (in $\mjup$)
%and final  (in $\msol$) masses, $L_1$
%the log of the luminosity (in units of $\lsun$) at 1 Myr,
$\mdot_{\rm burst}$ is
the accretion rate during the bursts (in $\msolyr$), $\dt_{\rm quiet}$
 the quiescent phase duration (in units of 10$^3$ yr)
and $N_{\rm burst}$
the total number of bursts. In all calculations, the burst duration varies between $\sim 75$ and $\sim$ 100 years. The label $a$ or $b$ indicates
whether the bursts start at the beginning of evolution (case $a$) or after a first phase of constant accretion ($\mdot = 10^{-5} \msolyr$ during $\dt = 10^{4}$ yr),
as suggested by Vorobyov \& Basu (2005) (case $b$).
\label{table2}}
\begin{tabular}{lccccccccc}
\tableline\tableline
case & $\minit$ & $m_{\rm f}$ &  $T_{\rm eff1}$ &$L_1$ & $\te^{\rm BCAH98}$ & $L^{\rm BCAH98}$ & $\mdot_{\rm burst}$ & $\dt_{\rm quiet}$ & $N_{\rm burst}$ \\
 \tableline
1a &1 & 0.1 & 3151 & -2.00 & 3001 & -1.16 & 10$^{-4}$ & 1 & 10 \\
1b &1 & 0.2 & 3373 & -1.25 & 3193 & -0.69 & 10$^{-4}$ & 1 & 10 \\
2b & 1 & 0.41 & 3650  &-0.90 & 3384 & -0.46 & 10$^{-4}$ & 1 & 30 \\
3b & 10 & 0.38 & 3466 & -1.05  &3335 & -0.43 & 10$^{-4}$ & 1 & 30 \\
4a & 10 & 0.05 & 2833 & -1.77  & 2844 & -1.73 & 10$^{-5}$ & 5 & 50 \\
5a & 50 & 1.35 & 4379 &  -0.16 & 4152 & 0.37 & 10$^{-4}$ & 1 & 150 \\
6a & 100 & 1.16 &4039 &  0.00 &3924 & 0.23 &  3 10$^{-4}$ & 1 & 40 \\
7a & 100 &1.67 & 4482 & 0.19 & 4317 & 0.53 & 3 10$^{-4}$ & 1 & 60 \\
\tableline
\end{tabular}
\end{center}
\end{table*}

\section{Discussion and conclusion}

The main
results presented in \S 3 can be summarized as follows: (i) for $\alpha=0$,
an accreting object
has a more compact structure at $\sim$ a few Myr, {\it i.e} a smaller radius and thus a smaller luminosity - looking older -, than the non-accreting counterpart of same mass and age;
(ii) if a fraction of the accretion energy is absorbed by the protostar interior ($\alpha \ne 0$), this extra energy source partly compensates or,  for $\alpha \simgr 0.2$ 
 (assuming $\epsilon$=1/2), even dominates the contraction
due to mass accretion,
leading in the latter case to a larger radius than
the non-accreting object of same mass and age.
Our results show that a scenario based on early stages of non-steady accretion, characterized by
short duration (a few 100 yr) episodes of vigorous accretion ($\mdot \sim 10^{-4} \msolyr$),
followed by longer quiescent ($\mdot \lesssim 10^{-5} \msolyr$) phases,
naturally produces a spread in the HR diagram at ages of $\sim$ a few Myr, providing non spherical accretion  (through disk or funnels) 
occurs on a variety of proto-star/BD initial masses, from a few $\mjup$ to a few tenths of $\msun$.  
%Objects undergoing such phases are denser, thus fainter, looking older, than their non accreting counterparts of same mass at an age of 1 Myr.
This scenario can easily produce a luminosity spread equivalent to an age spread of 
$\sim$ 10 Myr (for non-accreting objects), even though the objects are just 1 or so Myr old.  We find that such a spread can be obtained if only a negligible amount
of the accretion energy contributes to the contracting object's internal energy ($\alpha \ll 1$), {\it i.e} if most of the accreting kinetic
energy is radiated away. 
%In other words if the entropy of the accreted material is much lower than the one of the accreting object. 
%This is consistent with accretion through a viscous disk, with most of the accretion shock energy being radiated away before matter falls onto the protostar surface. 

Our scenario, however, cannot explain the observed population of very luminous low mass objects, which lie largely above the 1 Myr non-accreting isochrone, as displayed
in Figs. \ref{hrd1} and \ref{burst}. As discussed in \S \ref{subsection_test},
calculations with $\alpha \ne 0$ cannot maintain luminous objects at ages of $\sim$ Myr, with accretion rates consistent with the observed ones at this age,
because of the too short thermal timescale. 
%Or all these objects are phase, which seems statistically unlikely.
%a substantial deposit of accretion energy,  {\it i.e} calculations with $\alpha \ne 0$, provides evolutionary sequences which spend most of their early time at high luminosity
%and large radius. However, as soon as the phase of high accretion terminates,
%the object rapidly contracts and cannot remain at such high luminosity after
%$\sim$ 1 Myr of evolution. 
It thus seems difficult to explain the luminosity
of these objects with accretion.
A possible explanation is that they are significantly younger ($\ll $ 1 Myr) than the mean cluster age and experienced 
their episodic accretion phase quite recently.
%This interpretation, however, may face a timescale problem. Inferring for these objects a mass based
%on non accreting cooling tracks, their high luminosity and relatively low $\te$ yield
%a thermal timescale $\ll$ 1 Myr. These objects should thus contract
%on typical timescales of $\sim$ 10$^5$ yr. Therefore, the fraction of such objects in
%SFRs older than 1 Myr  should be at most of the order of $\sim$ 10\%, probably less. 
%In the Gatti et al. (2006) and Peterson et al. (2008) sample, the fraction of
%over-luminous objects is respectively 30\% and 23\%, which seems to be too large
%according to the previous argument. 
%If these observations are confirmed, 
%A more plausible explanation is that
%the mass of these objects is larger than previously estimated, implying a longer thermal timescale.
%As discussed in Chabrier et al. (2007), 
Another explanation, as discussed in Chabrier et al. (2007), is that fast rotation and/or the presence of a magnetic
field yields a smaller heat flux output, thus (i) a larger radius and (ii) a cooler $\te$,
 while barely affecting the luminosity, for a given mass. The net effect would be to shift the location of  $\sim$ 1 Myr old objects of a given mass
at cooler $\te$ for a given $L$, and thus 
on the right side of the 1 Myr non-accreting isochrone.
%or conversely a larger mass for a given $\te$. 
Note that this may apply as well to the other (hotter) objects, even though there is no need to invoke such a process to reproduce their luminosity. This suggestion should motivate observational determinations of the rotation velocity and the level of magnetic activity
of these over-luminous objects. The present analysis supports the conclusion
of Mohanty et al. (2009) on the origin of the $\te$ reversal 
in the young eclipsing binary brown dwarf 2M0535-05, which excludes
an explanation based on prior accretion.
At last, another explanation for the location of these objects in the HR diagram
 is a less reliable photometry. All the over-luminous objects in the Peterson et al. (2008) sample, for instance,
are located in a region of the Orion Molecular Cloud with very high nebulosity (D. Peterson, priv. comm.). 
%A careful re-analysis of these anomalously bright objects, depending on their location in the star forming regions,
%may help nailing down this issue. 

The suggestion, as explored in this Letter, that episodic accretion provides a plausible explanation for the observed spread in HR diagrams at ages of a few Myr (see \S \ref{intro}), is supported
by recent observations of protostars (Enoch et al. 2009; Evans et al. 2009).
Furthermore, our predicted significant spread in radius at ages of
$\sim$ Myr (see Fig. \ref{radius}) is consistent with the recently suggested
existence of such a spread in the ONC, based on the rotation periods and projected radial velocities of low mass objects (Jeffries 2007). Episodic accretion thus seems to provide several
matching pieces to the puzzle describing the early evolution of LMS and BDs. If our suggestion is correct, it has several drastic consequences: (1) what was interpreted as a significant age spread in SFRs or young clusters is essentially a spread in radius, thus in luminosity, for objects of comparable ages, (2) if young low-mass objects have experienced strong episodes of accretion during their embedded (class 0 to I) phase, their contraction proceeds very differently from a standard, constant $\te$, Hayashi track and still keep memories of these early episodes after about a few Myr, even if present accretion rates are negligible, (3) trying to infer the mass, thus an IMF, for young ($\lesssim Myr$ old) clusters from  mass-luminosity or mass-$\te$-Sp type  relationships based on non-accreting objects/models very likely leads to results of low significance, and inferring the IMF for such young associations seems to be elusive. 
At this stage, determination of the proto-star/BD {\it core} mass function (CMF) with submillimeter surveys will bring more robust information about the star formation process and the stellar mass spectrum (see e.g. Hennebelle \& Chabrier 2008, 2009). 
Finally, the present calculations show that the concept of a stellar birthline has no real significance, at least for low-mass objects, as the first appearance of these objects in a HR diagram is very random, due to the variety of prior accretion histories.

\acknowledgments
We thank D. Peterson and A. Natta for providing tables of their data,
K. Luhman, D. Peterson, S. Mohanty, A. Scholz and R. Jayawardhana for valuable discussions. I. B and G. C thank the astronomy department of the University of St Andrews for hospitality. 
%where part of this work was completed,
%for their warm hospitality. 
This work was
supported by the Constellation network MRTN-CT-2006-035890 and the french ANR %"Magnetic Protostars and Planets" 
MAPP project.


\begin{thebibliography}{}

%\bibitem[]{} Alibert, Y., Baraffe, I., Benz, W., Laughlin, G., Udry, S . 2009,
%"Structure formation in the Universe" , ed. G. Chabrier, Cambridge University
%Press, p.  378
\bibitem[]{} Ballestero-Paredes, J., Hartmann, L. 2007, RMxAA, 43, 123
\bibitem[]{baraffe98} Baraffe, I., Chabrier G., Allard F.,
%Hauschildt P.H., 1998, \aap, 337, 403
\bibitem[1997]{cb97} Chabrier, G., \& Baraffe, I. 1997, A\&A, 327, 1039
\bibitem[]{} Chabrier, G., Gallardo, J., Baraffe, I. 2007, \aap, 472, L17
%\bibitem[]{} Chabrier, G., Baraffe, I., Selsis, F., Barman, T., Hennebelle, P., Alibert, Y. 2007, Protostars and Planets V, eds. B. Reipurth, D. Jewitt, 
%and K. Keil, p. 623
%\bibitem[]{} Duch\^ene, G., M\'enard, F., Muzerolle, J., Mohanty, S. 2009, "Structure formation in the Universe", ed. G. Chabrier, Cambridge University Press
\bibitem[]{} Dunham, M.M., Crapsi, A., Evans, N.J. et al. 2008, \apjs, 179, 249
\bibitem[]{} Enoch, M.L., Evans, N.J., Sargent, A.I., Glenn, J. 2009, \apj, 692, 973  
\bibitem[]{} Evans, N.J., Dunham, M.M., Jorgensen, J.K. et al. 2009, \apjs, in press,
arXiv:0811.1059
%\bibitem[]{} Gallardo, J., Baraffe, I., Chabrier, G. 2008, proceedings of Cool Star 15, arXiv:0810.2931
\bibitem[]{} Gallardo, J., Baraffe, I., Chabrier, G. 2009, \aap, submitted
\bibitem[]{} Gatti, T., Testi, L., Natta, A., Randich, S., Muzerolle, J. 2006, \aap, 460, 547
\bibitem[]{} Gatti, T., Natta, A., Randich, S., Testi, L., Sacco, G. 2008, \aap, 423, 432
\bibitem[]{} Hartmann, L. 2001, \aj, 121, 1030
%\bibitem[]{} Hartmann, L., Kenyon, S. J. 1996, \araa, 34, 207
\bibitem{Hartmann} Hartmann, L., Cassen, P., Kenyon, S.J. 1997, \apj, 475, 770 
\bibitem[]{} Herczeg, G.J., Hillenbrand L. 2008, \apj, 681, 594
\bibitem[]{} Hennebelle, P. \& Chabrier, G. 2008 \apj, 684, 395
 \bibitem[]{} Hennebelle, P. \& Chabrier, G. 2009, \apj, in press
\bibitem[]{} Hillenbrand, L. 2008, IAU Symp. 258, eds. E Mamajek, D. Soderblom, R. Wyse
%\bibitem[]{} Hillenbrand, L.,  Bauermeister, A., White R.J. 2008, ASP Conf. Series, 384, 
ed. G. van Belle, p. 200
\bibitem[]{} Jeffries, R. D. 2007, \mnras, 381, 1169
\bibitem[]{} Kenyon, S.J., Hartmann, L., Strom, K.M., Strom, S.E. 1990, \aj, 99, 869
\bibitem[]{} Kenyon, S. J., Hartmann, L. 1995 \apj, 101, 117
%\bibitem[]{} Luhman, K., Joergens, V., Lada, C., Muzerolle, J., Pascucci, I., White, R. 2007, Protostars and Planets V, eds. B. Reipurth, D. Jewitt, 
%and K. Keil, p. 443
%\bibitem[]{}McKee, C., Ostriker, E.C. 2007, ARA\&A, 45, 565
% \bibitem[1984]{mercer} Mercer-Smith, J. A., Cameron, A. G., \& Epstein, R. I. 1984, ApJ, 279, 363
 \bibitem{Mohanty} Mohanty, S., Jayawardhana, R., Basri, G.  2005, \apj, 626, 498
 \bibitem[]{} Mohanty, S., Stassun, K.G., Mathieu, R. 2009, \apj, in press
\bibitem{Muzerolle} Muzerolle, J. et al.  2005, \apj,  625, 906 
% \bibitem{Muzerolle} Muzerolle, J.,  Luhman, K., Bricenyo, C., Hartmann, L., Calvet, N. 2005, \apj,  625, 906 
 \bibitem[]{} Myers, P. C. et al. 1998, \apj, 492, 703
 %\bibitem[]{} Myers, P. C., Adams, F.C., Chen, H., Schaff, E. 1998, \apj, 492, 703
%\bibitem{Natta} Natta, A., Testi, L.,  Randich, S.  2006, A\&A, 452, 245
 \bibitem[]{} Peterson, D.  et al. 2008, \apj, 685
  %\bibitem[]{} Peterson, D., Megeath, S.T., Luhman, K. et al. 2008, \apj, 685
 % \bibitem[]{} Prialnik, D., Livio M. 1985, \mnras, 216, 37
 % \bibitem[]{} Shu, F. H. 1977, \apj, 214, 488
 % \bibitem[]{} Siess, L., Forestini, M., Bertout, C. 1999, A\&A, 342, 480
\bibitem[]{} Siess, L. et al. 1997, A\&A, 326, 1001
%\bibitem[]{} Siess, L., Forestini, M., Bertout, C. 1997, A\&A, 326, 1001
%\bibitem{Stahler} Stahler, S. W.  1988, \apj, 332, 804
\bibitem[]{} Tassis, K., Mouschovias, T.C. 2005, \apj, 618, 769
\bibitem[]{} Vorobyov, E.I., Basu, S. 2005, \apj, 633, L137
%\bibitem[]{} Vorobyov, E.I., Basu, S. 2006, \apj, 650, 956
%\bibitem[1983]{whitworth06} Whitworth, A. P., \& Ward-Thompson, D. 2001, ApJ, 547, 317
%\bibitem[]{} White, R.J. et al. 2007,  PPV, eds. B. Reipurth, D. Jewitt, 
%and K. Keil, p.117
\bibitem[]{} Young, C. H., Evans, J.E. 2005, \apj, 627, 293
\bibitem[]{} Zhu, Z., Hartmann, L., Gammie, C. 2008, \apj, in press,
arXiv:0811.1762
\end{thebibliography}
\end{document}